\documentclass[aip,amsmath,amssymb,apl,reprint]{revtex4}
\usepackage{amssymb}
\usepackage{txfonts}
\usepackage{graphicx}

\begin{document}

\title{The methods to detect vacuum polarization by evanescent modes}

\author{Wei Li,$^{1,2,3}$ Jun Chen,$^{2,3}$ Gerard Nouet,$^{3}$, Liao-yao Chen,$^{4}$ and Xunya Jiang$^{1}$}
\email{xyjiang@mit.edu}

\address{1.State Key Laboratory of Functional Materials for Informatics, Shanghai Institute
of Microsystem and Information Technology, Chinese Academy of Sciences, Shanghai 200050,
China}
\address{2.Laboratoire de Recherche sur les Proprietes des Materiaux Nouveaux EA4257,
  IUT d'Alencon,Universite de Caen, 61250 Damigny, France}
\address{3.Centre de Recherche sur les Ions, les Materiaux et la Photonique, Caen 14050, France}
\address{4.Department of Optical Science and Engineering, Fudan University, Shanghai 200433, China}

\begin{abstract}
We propose the evanescent-mode-sensing methods to probe the
electrodynamics (QED) vacuum polarization. From our methods,
high-sensitivity can be achieved even though the external field is
much smaller than the Schwinger critical field and may be realizable
in contemporary experimental conditions. The methods are based on
the effect of phase change and time delay of evanescent wave which
is transmitted in QED vacuum. These methods can also be widely used
in sensitive probing of tiny dissipation in other fields.

\end{abstract}
 \maketitle

Vacuum is one of the most fundamental concepts in all quantum fields
of high-energy physics\cite{Perkins2000}, condensed-matter
physics\cite{Ashcroft1976}, statistical optics\cite{Goodman1985},
etc., since all excitations are from the vacuum and determined by
vacuum properties in some way. Modern vacuum concept is started from
quantum electrodynamics (QED), which describes the interaction
between light and matter (including vacuum), and has been widely and
continually studied both experimentally and
theoretically\cite{Feynman1985,DingKaplanPRL1989,
HeylJPA1996,RuffiniPR2009,GiesPRD2000,KingNP2010}. According to QED,
the vacuum becomes weakly anisotropic, dispersive, dissipative and
even nonlinear optical medium, when there is an external electric
field and its strength is approaching the Schwinger critical value
$E_c\simeq 10^{18}V/m$. In other words, the real and imaginary parts
of vacuum refractive index
 could deviate from unit and zero\cite{RuffiniPR2009,
DingKaplanPRL1989}, respectively. Physically, the deviation of the
imaginary part is mainly from the electron-positron pair generation.
However, the electron-positron pair generation, also generally
called as \emph{vacuum polarization} (VP)
processes\cite{RuffiniPR2009}, which is schematically shown in
Fig.1, has not been directly observed for over half century.

VP processes are very important to understand basic quantum
processes in many fields, \emph{e.g.} in condensed matter physics
where the ``electron-hole" pair generation is widely used in
calculating of electron self-energy and electron-phonon
interaction\cite{Ashcroft1976}. For QED vacuum, the obstacle to
observe VP is the very high critical electric field $E_c$ which is
beyond the contemporary technical limit. Therefore, it is natural to
wonder if we can find an approach to probe VP with external field
$E_{ext}$ much smaller than $E_c$.

\begin{figure}
{\includegraphics[width=0.25\textwidth]{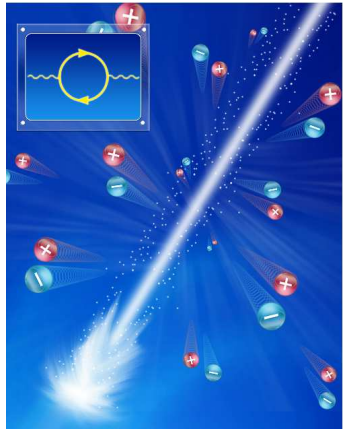}} \caption{The
schematic picture of vacuum polarization processes with
electron-positron pair generation, with which the vacuum becomes
dissipative and anisotropic. The insert is the Feynman diagram of
the vacuum polarization processes. The fermionic bold line
represents
 the coupling to all orders to the external electromagnetic field. }\label{vacuum}
\end{figure}

On the other hand, evanescent electromagnetic wave is intensively
studied recently because of its potential usage in the sensitive
detectors and other
directions\cite{HunspergerSV1985,BarwickNature2009,CarnigliaJOSA1971}.
In this work, we propose the evanescent-mode-sensing methods to
detect the QED VP, which is based on the measuring the phase change
and the time delay of \emph{evanescent waves} in the vacuum. We find
that the required external field could be one order weaker than
$E_c$, which may be realizable by contemporary experiments.

The idea of this work is from the thought of  ``dual roles" of
 real and imaginary parts of refractive index $n$ for the radiative waves or the
evanescent waves. Supposing a medium with complex dielectric
constant $n= n' + i n''$ where $n' \simeq 1$ and $ n''$ are
extremely small, our goal is to detect the very tiny change of $n'$
and $n''$. For radiative waves, $n'$ determines the real part of
wavevector $k \simeq n' \omega/c$, and we can easily measure the
phase change or group delay to detect the change of $n'$. So, it is
natural to choose the radiative wave as the probing light for
measuring the index real part change, as what have been done in the
experiments to detect the vacuum birefringence effect. On the other
hand, for radiative waves, very tiny $n''$ only causes an extremely
small intensity decay, which is very hard to measure in the limited
laboratory space. However, for the evanescent waves, the roles of
$n'$ and $n''$ are totally exchanged, i.e., $n'$ dominates the decay
rate, while the tiny $n''$ introduces a real part of wavevector in
the decay direction and causes a phase change which is much easier
to detect. Further more, we will demonstrate that the tiny imaginary
part $n''$ can also introduce the energy propagation whose energy
velocity $v_e \propto n''$ is extremely slow.
 Such a slow wave can be detected by measuring the delay time $\tau$
at a very short distance. The phase change $\Delta \phi$ and the
time delay $\tau$ can be used for sensitive detecting, especially
for the QED VP.

We would like to emphasize the mechanism difference between ours and
that in the previous work\cite{CarnigliaJOSA1971}, which is based on
the systems with at least ``two interfaces" (such as a slab). Such
``two-interfaces" structure will generate both evanescent modes
$\exp(\kappa x)$ and $\exp(-\kappa x)$ at the same time, and such
two evanescent modes can carry electromagnetic energy
current\cite{LiuZiAPL2010}, which is the essence of ``tunneling
effect". So, even the dissipation of material could be neglected
(dissipation is truly neglected in that work), the propagation of
electromagnetic energy is still available based on that tunneling
mechanism. However, in our model, the mechanism is totally
different, since there is only one single interface in our model
(Fig.2, the details can be seen in the following). The obvious
evidence is that, if without dissipation in our ``single-interface"
model, there will be no the energy current at all in the
no-dissipation medium\cite{JAKong1990}, then, the phase change is
zero and the energy delay time makes no sense. But, we know there is
tiny dissipation in the QED vacuum because of VP, so that the phase
change and the energy delay time in our model are not from tunneling
effect, but purely from QED vacuum dissipation.

Here we note that,  because the probing light is much weaker than
the external field in our study, the nonlinear effect is negligible.
Since it is a linear problem, all dynamical processes, such as the
propagation of envelope fluctuation of the transmitted evanescent
wave, can be solved by sum of multi-frequency components. Based on
the linear property, we can use Green's function \cite{JAKong1990}
of multi-frequency components to obtain the strict numerical
results, which can be compared with our analytical ones of dynamical
process of evanescent wave.

\begin{figure}
{\includegraphics[width=0.45\textwidth]{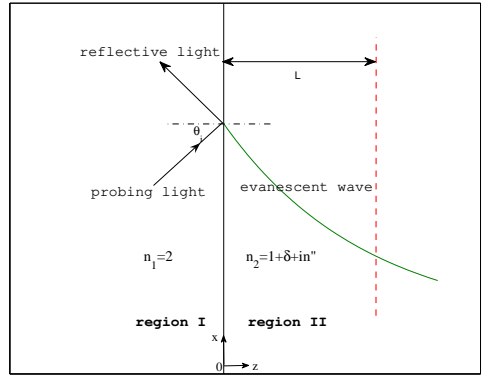}} \caption{The
schematic diagram of our model.}\label{modelslab}
\end{figure}

Our model is schematically shown in Fig.\ref{modelslab}, based on
the total internal reflection (TIR) at the interface between a
dielectric media $n_1$ (region $I$) and vacuum (region $II$). When
the incident angle $\theta_i > \theta_c = arcsin({1}/{n_1})$, the
TIR will occur and the transmitted wave in the vacuum is the
evanescent wave. We choose $\theta_i$ is a little larger than
$\theta_c$ to make sure that almost all frequency components are
totally reflected when the incidence is the slowly-varying
quasi-monochromatic wave. An interferometer or a photon detector is
set at distance $L$ from the interface so that the phase information
and intensity change can be detected.

The time-dependent Maxwell equations are given by
$\nabla\times\bold{E}=-\mu(z)\mu_0\partial\bold{H}/\partial t$ and
$\nabla\times\bold{H}=\epsilon(z)\epsilon_0\partial\bold{E}/\partial
t$, where $\epsilon(z)$ and $\mu(z)$ are the relative permittivity
and the relative permeability, respectively, and
$c=1/\sqrt{\epsilon_0\mu_0}$. To obtain the concrete results, the
system parameters are chosen as following,
 the incident angle $\theta_i=0.1667\pi$, the refractive index of region
$I$ $n_1=\sqrt{\epsilon_1}=2$,
 and the vacuum refractive index of region
$II$ $n_2=\sqrt{\epsilon_2\mu_2}=1+\delta+in''$, where $\delta << 1
$ and $n'' << 1 $ are the real and imaginary index deviations of
vacuum, because of VP processes caused by strong external field. If
the incident probing light is a plane wave, the transmitted wave in
the vacuum region can be generally written in the form $E(x,z,t)=E
exp(ik_z z +ik_{\parallel} r_{\parallel}-i\omega t)$, where
$k_{\parallel}=n_1 \sin \theta_i \omega/c$ and $k_z=\sqrt{(n_2
\omega/c)^2-k_{\parallel}^2}$ are the wave vectors parallel and
perpendicular to the interface. For the evanescent wave, $k_z$ is
described as:
\begin{equation}\label{kz}
    k_z= i\sqrt{(n_1\sin\theta_i)^2-(1+\delta)^2}\frac{\omega}{c}
+\frac{n''}{\sqrt{(n_1\sin\theta_i)^2-(1+\delta)^2}}\frac{\omega}{c}.
\end{equation}
The physical meaning of $k_z$ is very clear that the imaginary part
$Im(k_z)=\kappa_z$ corresponds to the exponential decay of the
field, and the real part $Re(k_z)$ causes \emph{a phase change
because of VP}. The phase change at distance $z=L$ is

\begin{equation}\label{phasechange}
\Delta \phi=Re(k_z)L \propto n'' L
\end{equation}
which could be measured by interferometers\cite{Lawall2000}.

\begin{figure}
{\includegraphics[width=0.45\textwidth]{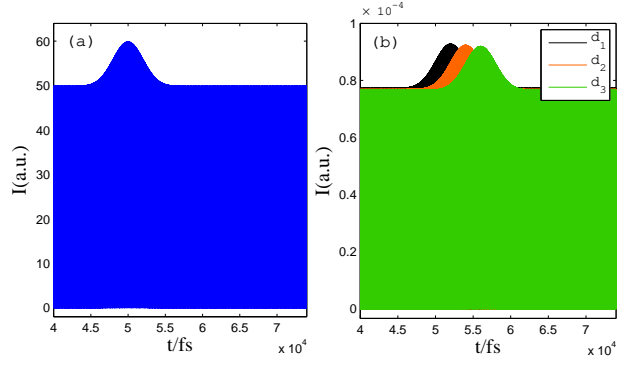}} \caption{The
irradiance of light $I$ versus time $t$. (a)The incident probing
light at the interface; (b)The transmitted evanescent wave in region
\emph{II}. The black one, the red one and the green one show the
evanescent wave at the distance from the interface
$d_1=0.1\lambda_0$, $d_2=0.2\lambda_0$ and $d_3=0.3\lambda_0$,
respectively, where $\lambda_0=600nm$ is the wavelength of the
center frequency component of the probing light in the classical
vacuum.}\label{delaytime2}
\end{figure}

Besides the phase change $\Delta \phi$, with the same model as shown
in Fig.\ref{modelslab}, there is another way to detect the tiny
$n''$ by \emph{measuring the time delay of
irradiance}\cite{SuXuJB2003} \emph{fluctuation of the evanescent
wave}. The physical process can be explained in the following way.
First, we suppose that the incident probing light is not a plane
wave anymore, but with a slow intensity fluctuation which is
proportional to the irradiance fluctuation, as shown in
Fig.\ref{delaytime2}(a), then the question is ``What will happen for
the evanescent wave in region \emph{II} ?" From the strict Green's
function method with physical dissipation and dispersion, it is
found that the fluctuation will propagate on the evanescent wave
from the interface to far away, as shown in Fig.\ref{delaytime2}(b).
So, we can measure the time delay $\tau$ of the fluctuation
propagation on the evanescent wave to detect the VP effect. The
propagation speed of irradiance fluctuation can be obtained by the
energy velocity $v_e$ which  is defined as: $v_e=|\vec S_z| / W$,
where $\vec S_z =\frac{1}{2}Re\left(E \times H^{*} \right)|_z$ is
the averaged Poynting vector along $z$ direction, and
$W\simeq\frac{1}{4}(\epsilon_0|E|^2+\mu_0|B|^2)$ is the local energy
density of the electromagnetic wave. In our model, one can obtain
the energy velocity as:
\begin{equation}\label{vez}
    v_{e}=\chi\cdot n''
\end{equation}
with $\chi\simeq
c/\left[(n_1\sin\theta_i)^2\sqrt{(n_1\sin\theta_i)^2-(1+\delta)^2}\right]$,
when the dissipation and dispersion are very weak\cite{generalcase}.
The physical meaning of $v_e$ can be understood as the
``propagation" speed of electromagnetic wave irradiance fluctuation
of the evanescent wave,  which can be measured by the irradiance
measurement technology\cite{SuXuJB2003}.  Here we note that with
tiny $n''$ the expected energy velocity is much smaller than $c$,
and hence causality is not violated.

Hence, experimentally the time delay  $\tau$ of the irradiance
fluctuation at distance $L$ can be measured:
\begin{equation}\label{tau}
\tau=L /v_e \propto 1/n''.
\end{equation}
Since it is \emph{near field} phenomenon, the detecting should be
very near the interface. For the VP effect, since $n''$ is extremely
small, the ``propagation" speed of the irradiance fluctuation is so
slow that $\tau$ is long enough for detecting even in a very short
distance $L$, i.g. $\tau$ gets to peco-second level when the
distance is one tenth of the wavelength $L = 60nm$.

Therefore, either the phase change $\Delta \phi$ or the time delay
$\tau$  are very
 sensitive for $n''$, and the evanescent wave is a good candidate to probe
the VP effect. Here, we note that the famous Kramers-Kronig
relations, which shows the confinement of causality limit, still fit
for QED vacuum\cite{RuffiniPR2009}. Hence, the direct observation of
imaginary part of vacuum index also confirms the dispersive property
of QED vacuum.

\begin{figure}
{\includegraphics[width=0.45\textwidth]{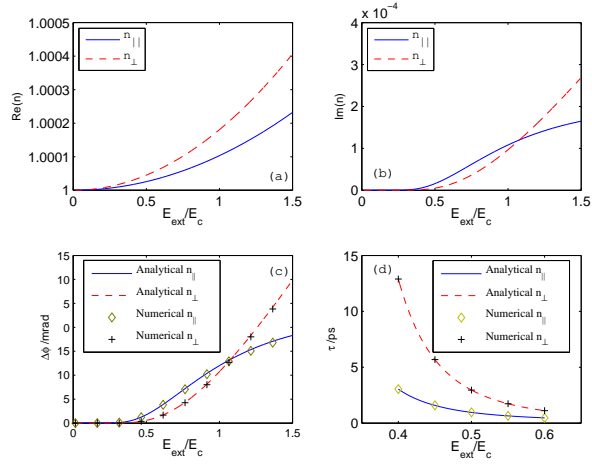}} \caption{The
refractive index, the phase change and time delay in the QED vacuum
with an external electric field $E_{ext}$. (a)The real part of
$n_{\parallel}$ and $n_{\bot}$ versus the external electric field
strength; (b)The imaginary part of $n_{\parallel}$ and $n_{\bot}$
versus the external electric field strength; (c)The phase change
versus the external electric field strength when the distance from
the interface $L_p$ is $6\mu m$; (d)The time delay versus the
external electric field strength when the distance from the
interface $L_{\tau}$ is $60nm$. Both the results from theory and
from Green's function are shown in (c) and (d).}\label{pvacuumdelay}
\end{figure}

Next, we will quantitatively study the detect of QED VP by our
methods. Supposing that an external homogeneous constant electric
field $E_{ext}$, which is perpendicular to the $xz$ plane and
smaller than the Schwinger critical electric field $E_c$, is applied
to the vacuum (region $II$) only, as shown in Fig.\ref{modelslab},
then, the optical properties of the vacuum can be described by the
Euler-Heisenberg Lagrangian
$L_{eff}$\cite{RuffiniPR2009,GiesPRD2000}. Physically, the imaginary
part of Euler-Heisenberg Lagrangian $L_{eff}$ is related to the
imaginary part of VP operator, and therefore corresponds to the
\textit{electron-positron pair generation}. This result of QED  is
well justified not only at zero-temperature but also at finite
temperature cases\cite{GiesPRD2000}. Consequently, the vacuum
refractive index can be deduced from the Lagrangian
$L_{eff}$\cite{RuffiniPR2009,GiesPRD2000,HeylJPA1996}.

In our model, the contribution of the transmitted evanescent wave to
$L_{eff}$ is negligible, since its electric field is much weaker
than $E_{ext}$, and furthermore, the external magnetic field is
supposed to be zero, thus the vacuum refractive index is determined
only by the external homogeneous constant electric field $E_{ext}$.
We use $n_{\parallel}$ and $n_{\perp}$ to refer the effective
refractive index of vacuum when the electric field of probing light
are parallel and perpendicular to the field $E_{ext}$, respectively.
$n_{\parallel}$ and $n_{\perp}$ can be obtained from the reference
\cite{HeylJPA1996}:

\begin{equation}\label{nparal}
 n_{\parallel}=1+\frac{2\alpha}{45\pi}y^2+i\cdot\frac{\alpha}{4\pi}
 \sum_{n=1}^{\infty}\left(\frac{\pi}{y^2}+\frac{1}{n}
 \frac{1}{y}\right)\exp(-n\pi/y),
\end{equation}
and

\begin{equation}\label{nperp}
 n_{\perp}=1+\frac{7\alpha}{90\pi}y^2+i\cdot\frac{\alpha}{4\pi}\sum_{n=1}^{\infty}
 \left(\frac{2}{3}\pi+
 \frac{1}{n}\frac{1}{y}+  \frac{1}{n^2}\frac{2}{\pi}\right)\exp(-n\pi/y),
\end{equation}
where $y=|E_{ext}|/E_c$, and $\alpha\simeq 1/137$ is the
fine-structure constant. Therefore we have
$\delta=Re\left(n_{\parallel(\perp)}\right)-1$,
$n''=Im\left(n_{\parallel(\perp)}\right)$ for $n_{\parallel}$ and
$n_{\perp}$ when we solve the equations such as Eq.(\ref{kz}) in
this letter.

The parameters of our model in Fig.\ref{modelslab} are chosen as
following. The wavelength of probing light is $\lambda_0=600nm$, the
dielectric constant in the region $I$ is  $\varepsilon=4$, and the
incident angle is $\theta_{inc}=0.1667\pi$, which is a little larger
than critical angle $\theta_c=\pi/6$ of TIR, so that the field in
vacuum is evanescent. The distance $L$ for the phase detecting is
$L_p=6 \mu m =10 \times \lambda_0$, while for the time delay
detecting is $L_{\tau}=60nm=0.1 \times \lambda_0$, respectively. The
QED theoretical results of real and imaginary part of
$n_{\parallel}$ and $n_{\perp}$ are shown in
Fig.\ref{pvacuumdelay}(a) and Fig.\ref{pvacuumdelay}(b),
respectively. Bring these results into Eq.(\ref{phasechange}) and
Eq.(\ref{tau}), the phase change $\Delta \phi$ and the delayed time
$\tau$  can be obtained, which are shown in Fig\ref{pvacuumdelay}(c)
and Fig.\ref{pvacuumdelay}(d), respectively. Numerically, the phase
change with plane wave incidence and the time delay of local
amplitude maximum are calculated by Green's function method,  which
are also shown in Fig.\ref{pvacuumdelay}(c) and
Fig.\ref{pvacuumdelay} (d). Comparing the analytical results from
Eq.(\ref{phasechange}) and Eq.(\ref{vez}) and numerical results,  we
can find that they agree with each other very well.

Next, we will analyze the possibility to observe the VP effect in
experimental conditions. The recent experimental
advances\cite{TESLATechnical} have raised hopes that lasers may
achieve fields just one or two orders of magnitude below the
Schwinger critical field strength. In this case $E_{ext} \sim 0.1
E_c$, from our numerical and analytical results in
Fig.\ref{pvacuumdelay}, we can see the $\Delta\phi$ can get to $
\sim 10^{-1} mrad$ order, which are in measuring limit of
contemporary interferometer \cite{Lawall2000}. Very recently, it is
supposed that the electric field $E$ could be effectively amplified
4 times larger by coherent constructive interference of laser
beams\cite{KingNP2010}. If $E_{ext}$ can get to $ 0.5 E_c$ by this
method, not only $\Delta \phi$ can be one order larger, but also the
delay time $\tau$ can get to sub peco-second level and  may be
measured by contemporary photon detectors.

In conclusion, we have investigated the evanescent-mode-sensing
methods to directly detect the QED VP based on evanescent wave.
Theoretically, we clearly demonstrate that the imaginary part of QED
vacuum index, caused by QED VP processes, can generate a phase
$\Delta \phi$ and a time delay $\tau$ of irradiance fluctuation
propagation on the evanescent wave. From Green's function method, we
obtain the numerical results of $\Delta \phi$ and $\tau$, which
agree with our analytical ones very well. The possibility to
directly observe the effects of VP based on evanescent wave is
discussed, and it is found that the required external electric field
could be much smaller than the Schwinger critical field and maybe
realizable in contemporary experiments. Our methods can also be used
in other extremely sensitive detections.

\textit{Acknowledgement}. This work is supported by the NSFC (Grant
No. 11004212, 10704080, 60877067 and 60938004), the STCSM (Grant No.
08dj1400303 and 11ZR1443800).


\begin{thebibliography} {s1}
\bibitem{generalcase}In our model, for the general case, we can obtain: $v_e=(1+\delta)^2\beta\cos(\alpha/2)+2n''\beta\sin(\alpha/2)$,
where $\alpha=Arg[n_2^2-n_1^2\sin^2\theta_i]$ is the principle value
of the argument, and
$\beta=\sqrt[4]{(2n'')^2+[n_1^2\sin^2\theta_i-(1+\delta)^2]^2}\cdot(1+\delta)^2c/[(1+\delta)^4\sin^2\theta_i+(2n'')^2\sin^2\theta_i]$.
\end{thebibliography}
\end{document}